\documentclass[amsmath,amsfonts,prl,aps,twocolumn,twoside,showpacs]{revtex4}
\usepackage{epsfig,verbatim}
\usepackage{graphicx}
\usepackage{color} 
\usepackage[caption=false]{subfig}

\newcommand{\be}{\begin{equation}}
\newcommand{\ee}{\end{equation}}
\newcommand{\ba}{\begin{array}}
\newcommand{\ea}{\end{array}}
\newcommand{\bea}{\begin{eqnarray}}
\newcommand{\eea}{\end{eqnarray}}

\newcommand{\trace}{\mathop{\mathrm{Tr}}\nolimits}

\newcommand{\ket}[1]{| #1 \rangle}
\newcommand{\bra}[1]{ \langle #1 |}

\renewcommand{\section}[1]{{\em #1}---}

\begin{document}
\title{Maximum Likelihood, Minimum Effort}

\author{John A. Smolin}\email{smolin@watson.ibm.com}
\affiliation{IBM T.J. Watson Research Center, Yorktown Heights, NY 10598, USA}

\author{Jay M. Gambetta}
\affiliation{IBM T.J. Watson Research Center, Yorktown Heights, NY 10598, USA}

\author{Graeme Smith}\email{gsbsmith@gmail.com}
\affiliation{IBM T.J. Watson Research Center, Yorktown Heights, NY 10598, USA}

\date{\today}

\begin{abstract}
  We provide an efficient method for computing the maximum likelihood
  mixed quantum state (with density matrix $\rho$) given a set of
  measurement outcome in a complete orthonormal operator basis subject
  to Gaussian noise.  Our method works by first changing basis
  yielding a candidate density matrix $\mu$ which may have nonphysical
  (negative) eigenvalues, and then finding the nearest physical state
  under the 2-norm.  Our algorithm takes at worst $O(d^4)$ for the
  basis change plus $O(d^3)$ for finding $\rho$ where $d$ is the
  dimension of the quantum state.  In the special case where the
  measurement basis is strings of Pauli operators, the basis change
  takes only $O(d^3)$ as well.  The workhorse 
  of the algorithm is a new linear-time method for finding the closest
  probability distribution (in Euclidean distance) to a set of real
  numbers summing to one.
\end{abstract}
\maketitle

{\em 
The race is not always to the swift nor the battle to the strong,
but that's the way to bet}---Damon Runyon\\



As scientists, we are faced again and again with the problem of determining
from imperfect data what ``really happened'' in an experiment.  Generally,
the more data one has the better the reconstruction of the true events.  Even
so, the view remains imperfect, so one typically tries only to determine
what was the 
event most likely to have led to the observed data.
When quantum mechanics is considered, the situation is harder on the
experimentalist, since even the results of perfectly performed measurements
may have probabilistic results.


Nevertheless, one can still determine the quantum state of a system by
performing many experiments on identically prepared systems and
building up good statistics on the outcomes.  If the set of
experiments is {\em informationally complete} then the mixed state
density matrix describing the system can be determined. This is called
quantum state tomography \cite{raymer,tomography,other}.  A
completely determination of the state would require an infinite number
of perfect measurements, which are unobtainable, so instead we
concentrate on finding the maximum likelihood state ({\em cf.}
\cite{hradil}).

We consider an informationally complete set of measurements, each
performed many times on identically prepared quantum systems.  From
the experimental outcomes, we would like to determine the quantum
state that gives the observed results with highest probability.  This
can be a computationally intensive task.  For the two qubit
experiments of \cite{Jerry}, conventional ML solving took more time than
the experiments themselves.  And it has been reported in \cite{steve} that
the ML
reconstruction for 8 qubits in \cite{Haffner} took {\em weeks} of computation.
Our main
result is a fast algorithm for reconstructing this state when the
noise is Gaussian.  For 8 qubits our algorithm runs in {\em seconds}.

The rest of the paper is organized as follows: First, we show that the
ML state reconstruction problem with Gaussian noise is equivalent to
a least-squares minimization problem on quantum states.  Next, we prove
that the minimum takes a particularly simple form.  Finally, we give a
fast algorithm that finds this minimum explicitly and benchmark it
against direct minimization.

\section{Reduction to density matrix minimization}
Observables in quantum mechanics are
Hermitian operators with thhe expectation value of a Hermitian operator
$\sigma$ applied to a mixed state $\rho$ given by $\trace (\sigma
\rho)$.  We can represent the result of an imperfect measurement of
such an expectation subject to additive Gaussian noise with variance $v$
as a probability density function 
$p(m|\rho)=\frac{1}{\sqrt{2 \pi v}}e^{-[m-\trace{(\sigma \rho)}]^2/(2 v)}$.

Given an orthonormal Hermitian operator basis
$\{\sigma_i\}_{i=1}^{d^2}$ on $d \times d$ matrices (with
$\trace[\sigma_i \sigma_j]=d \delta_{ij}$), and a particular
set of measured values $m_{ij}$ corresponding to $j$th measurement
result of expectation value $\sigma_i$ applied to the ``true state''
$\rho_0$, we want to find the mixed state
$\rho$, a trace 1 Hermitian matrix with only nonnegative eigenvalues,
maximizing the likelihood function
\begin{equation}
{\cal L}=\prod_{ij} p^{(i)}(m_{ij}|\rho)=\prod_{ij} 
\frac{1}{\sqrt{2 \pi v}}e^{-[m_{ij}-\trace{(\sigma_i \rho)}]^2/(2 v)}
\label{calL}
\end{equation} 
or
\begin{equation}
{\cal L}=\prod_i(\frac{1}{\sqrt{2 \pi v}})^n e^{-n[m_i-\trace{(\sigma_i \rho)}]^2/(2 v)}.
\end{equation}
Here $n$ is the number of measurement results for the expectation of $\sigma_i$
and $m_i =\sum_{j=1}^n m_{ij}/n$ is the average value of those results.
The same $\rho$ that maximizes ${\cal L}$ will minimize the 
log likelihood function
\begin{equation}
{\cal L}_{\log}=\sum_i [m_i-\trace{(\sigma_i \rho)}]^2\ .
\end{equation}

Working in the operator basis of the $\{\sigma_i\}$s is not
convenient, but fortunately the distance is just the Hilbert-Schmidt,
or 2-norm, which is basis independent.  We show this here for
completeness:

\noindent {\bf Lemma}:  $\sum_i (m_i-r_i)^2=||\mu-\rho||_2^2/d$
where $m_i=\trace[\sigma_i \mu]$ and $r_i=\trace[\sigma_i \rho]$.

{\em Proof:}
\begin{eqnarray}
||\mu-\rho||_2^2&=&\trace(\mu-\rho)^2\\
\nonumber&=&\trace[(\sum_i (m_i-\trace \rho \sigma_i)\sigma_i)^2]\\
\nonumber&=&\sum_{ij}\trace[(m_i-\trace \rho \sigma_i)\sigma_i (m_j-\trace \rho \sigma_j)\sigma_j]\\
\nonumber&=&\sum_{ij}(m_i-\trace \rho \sigma_i)(m_j-\trace \rho \sigma_j)\trace[\sigma_i\sigma_j]\\
\nonumber&=&d \sum_i(m_i-\trace \rho \sigma_i)^2
\end{eqnarray}
$\Box$

The matrix $\mu=(1/d) \sum_i m_i \sigma_i$ can be thought of as the
experimentally noisy view of the density matrix $\rho_0$.  Note that
it is trace one by construction, but may have negative eigenvalues.
Calculation of $\mu$ from the $m_i$s is a change of operator basis,
and in general requires time $O(d^4)$ (there are $d^2$ values of $i$
and each $\sigma_i$ is a $d \times d$ matrix).  This will actually be
the limiting step in our overall algorithm, as all other steps with be
$O(d^3)$ or better. In many cases of interest, however, the operator
basis change can be done more quickly.  This will be true whenever the
matrices representing the $\sigma_i$s in the canonical basis are
sparse.  In particular, if the $\sigma_i$s are tensor products of the
Pauli matrices
\begin{equation}
\nonumber
\{\sigma_0,\sigma_1,\sigma_2,\sigma_3\}=
\{ {1\ 0 \choose 0\ 1},{0\ 1 \choose 1\ 0},{0\ -\!i \choose i\ \ \ \ 0},
{1\ \ \ \  0 \choose 0\ -\!1} \}
\end{equation}
on $n$ qubits, so that $d=2^n$,
each $\sigma_i$ has only $d$ nonzero elements and the change of basis
can be carried out in $O(d^3)$ steps.

Now, after the change of basis from the $m_i$s to $\mu$, our original
maximum-likelihood problem has been transformed into the following:

\noindent {\bf Subproblem 1}:
Given a trace-one
Hermitian matrix $\mu$,
find the closest density matrix $\rho$ (a
trace-one Hermitian matrix with only nonnegative eigenvalues) under
the 2-norm:
\begin{equation}
||\mu-\rho||_2^2 =\trace[(\mu-\rho)^2]=\sum_{ij} |\mu_{ij}-\rho_{ij}|^2
\label{norm}
\end{equation}

This is immediately familiar as a least squares minimization problem,
for which standard minimizer packages are well suited.  Indeed, this
is how the problem is 
often solved in practice.  Unfortunately, finding the solution can be
computationally intensive.  In standard state reconstruction
algorithms, this is the most expensive step by far (see
Fig.~\ref{figure}). 

\section{Simple form for the minimum}
To improve matters, 
since the 2-norm is basis independent, we can work in the eigenbasis of 
$\mu$.  Then we observe that the optimum $\rho$ is diagonal
in this basis as well.  This is immediate from the form of (\ref{norm}),
where any off diagonal terms could only contribute positive amounts to the sum.
Thus, the problem reduces to finding the eigenvectors
and eigenvalues of $\mu$ and picking the $d$ non-negative eigenvalues
for $\rho$ that minimize (\ref{norm}).  Eigensystem solutions are
$O(d^3)$ and good packages exist \cite{LAPACK}.

We are left with a minimization over $d$ variables, effectively the
square root of the difficulty of the original problem.  Call the
eigenvalues of $\mu,\rho$  $\mu_i,\lambda_i$ and arrange them such that $\mu_i \ge \mu_{i+1}$.  We now want to minimize $\sum_i
(\lambda_i-\mu_i)^2$ such that $\sum_i \mu_i = \sum_i \lambda_i =1$ and $\lambda_i\ge 0$.
Using the method of Lagrange multipliers to impose this constraint,
and writing $\lambda_i=x_i^2$ to enforce non-negativity of $\lambda_i$
we write the objective function
\begin{equation}
\Lambda=\sum_i (x_i^2-\mu_i)^2 - L (\sum_i x_i^2 -1 ) \ .
\end{equation}
Differentiating with respect to $x_i$ we have
\begin{equation}
\frac{\partial \Lambda}{\partial x_i}= 4 (x_i^2 -\mu_i) x_i- 2 L x_i =0\ .
\end{equation}
This equation has two solutions, either $x_i=0$ or 
\begin{equation} x_i^2=L/2 + \mu_i\ .
\label{xi}
\end{equation} 
Note that $L$ doesn't depend on $i$ so each $\lambda_i=x_i^2$ is either set to
zero or given by $\mu_i$ plus the very same number.  To evaluate
$L$ we must pick a set ${\bf s}=\{i {\rm\ such\ that\ } x_i\ne 0\}$.  
Then,
summing (\ref{xi}) over $i$ we have 
$1=\sum_i x_i^2 = |{\bf s}| L/2 + \sum_{i \in s}\mu_i$ or
\begin{equation}
L/2 =\frac{1}{|{\bf s}|}\sum_{i\notin {\bf s}} \mu_i\ 
\end{equation}
and
\begin{equation}
\Lambda=\frac{1}{|{\bf s}|}(\sum_{i \notin {\bf s}} \mu_i)^2 +
\sum_{i \notin {\bf s}} \mu_i^2\ .
\end{equation}

\noindent {\bf Lemma}:
Consider an $i$ and $j$ with $\mu_i > \mu_j$ with $i \in {\bf s}$ and
$j \notin {\bf s}$.  Then $\Lambda_i$, the distance function for this case, is
always less than or equal to  $\Lambda_j$, the distance function for a new set 
${\bf s'}={\bf s} + \{j\} - \{i\} $.
 
\noindent {\em Proof:}
We can write
\begin{equation}
\Lambda_i = |{\bf s}|\left(\frac{L}{2}\right)^2 + \sum_{k\notin {\bf s}}\mu_k^2
\label{lambdai}
\end{equation} and for the case with $j\in {\bf s}$ and $i\notin {\bf s}$
\begin{equation}
\Lambda_j = |{\bf s}|\left(\frac{L}{2} +\frac{\mu_i-\mu_j}{|{\bf s}|}\right)^2  -\mu_j^2+ \mu_i^2 +\sum_{k\notin s}\mu_k^2\ .
\end{equation}
If we had $\Lambda_j<\Lambda_i$ this would imply
\begin{equation}
L<\frac{\mu_j-\mu_i -|{\bf s}|(\mu_i+\mu_j)}{|{\bf s}|}\ .
\end{equation}
Then we would have 
\begin{equation}
\lambda_j=\mu_j+\frac{L}{2}+\frac{\mu_i-\mu_j}{|{\bf s}|}
<\frac{(|{\bf s}|-1)(\mu_j-\mu_i)}{2 |{\bf s}|}\le 0
\end{equation}
because $\mu_i \ge \mu_j$.  But 
$\lambda_j$ must be nonnegative, therefore 
$\Lambda$ is never decreased by moving $i$ into and $j$ out of ${\bf s}$.\\
$\Box$

The lemma tells us that all the $\lambda_i$s that are zero are together at
the end, matching up with the smallest $\mu_i$s.
Thus, rather than the $2^d$ possible choices for ${\bf s}$ we need only
decide where to put the break between zero and nonzero $\lambda_i$s, for
which there are only $d$ choices.  

Next, we show that the choice of ${\bf s}$ should be the largest set 
satisfying the constraint that all the $\lambda_i$s are nonnegative.
Starting from Eq. (\ref{lambdai}), we imagine removing some element $j$ from
from ${\bf s}$.  Then 
\begin{equation}
\Lambda'=\frac{1}{|{\bf s}|-1} \left(|{\bf s}| \frac{L}{2} + \mu_j\right)^2 + \mu_j^2
+\sum_{k \notin {\bf s}} \mu_k^2
\end{equation}
and
\begin{eqnarray}
\nonumber\Lambda'-\Lambda_i&=&\frac{1}{|{\bf s}|-1} \left(|{\bf s}| \frac{L}{2} + \mu_j\right)^2 + \mu_j^2
-|{\bf s}| \left(\frac{L}{2}\right)^2 \\
\nonumber&=&\frac{|{\bf s}|}{|{\bf s}|-1}
\left((\frac{L}{2})^2 + 2 \mu_j \frac{L}{2} +  \mu_j^2\right)\\
&=&\frac{|{\bf s}|}{|{\bf s}|-1} \left(\frac{L}{2}+\mu_j \right)^2 \ge 0\ .
\end{eqnarray}
In other words, setting any more of the $\lambda_i$s to zero than necessary increases the
distance function.  We are now ready to give an algorithm for Subproblem 1.

\section{Fast algorithm for Subproblem 1}

\begin{enumerate}
\item Calculate the eigenvalues and eigenvectors of $\mu$.  
Arrange the eigenvalues is order from largest to smallest.
Call these $\mu_i, \ket{\mu_i}$, $1\le i \le d$.
\item Let $i=d$ and set an accumulator $a=0$.
\item If $\mu_i+a/i$ is non-negative, go on to step 4.  
Otherwise, set $\lambda_i=0$ and add $\mu_i$ to $a$.
Reduce $i$ by 1 and repeat step 3.
\item Set $\lambda_j=\mu_j+a/i$ for all $j\le i$.
\item Construct $\rho = \sum_i \lambda_i \ket{\lambda_i}\!\bra{\lambda_i}$.
\end{enumerate}

\begin{figure}
\subfloat[]{\includegraphics[width=4.1cm]{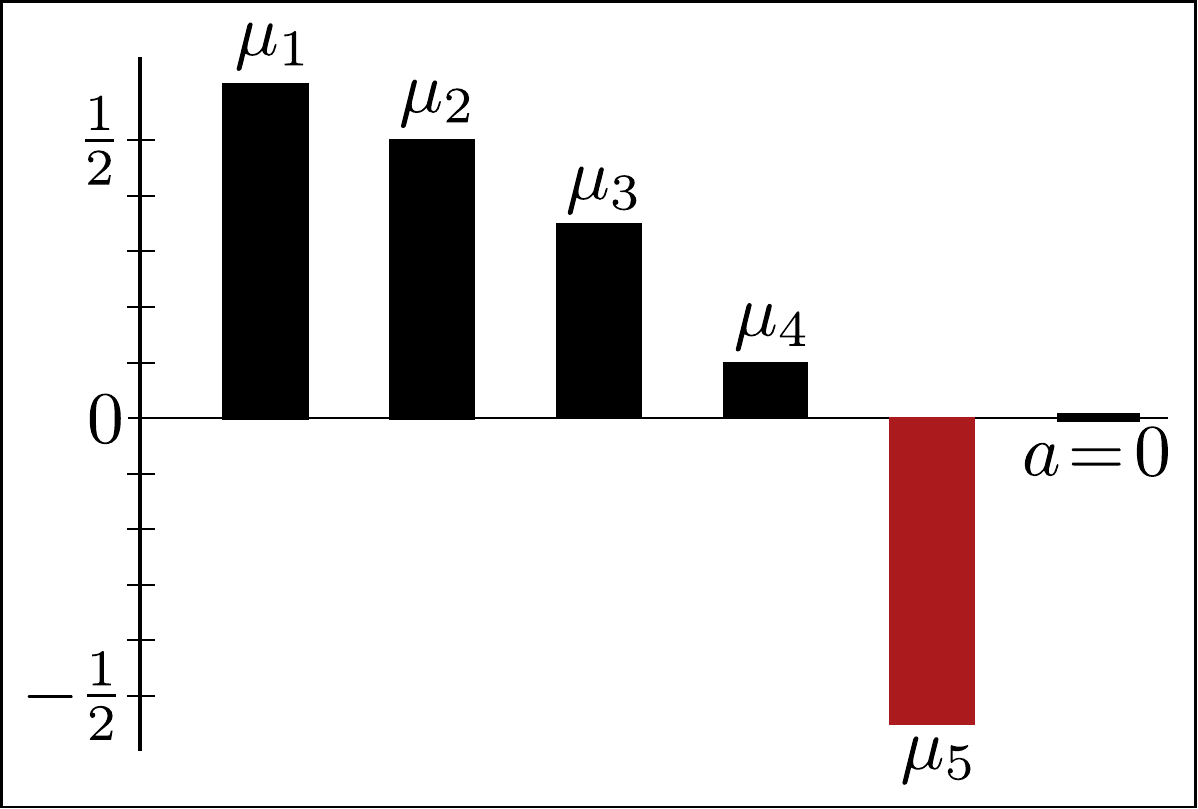}}
\ \ \ \subfloat[]{\includegraphics[width=4.1cm]{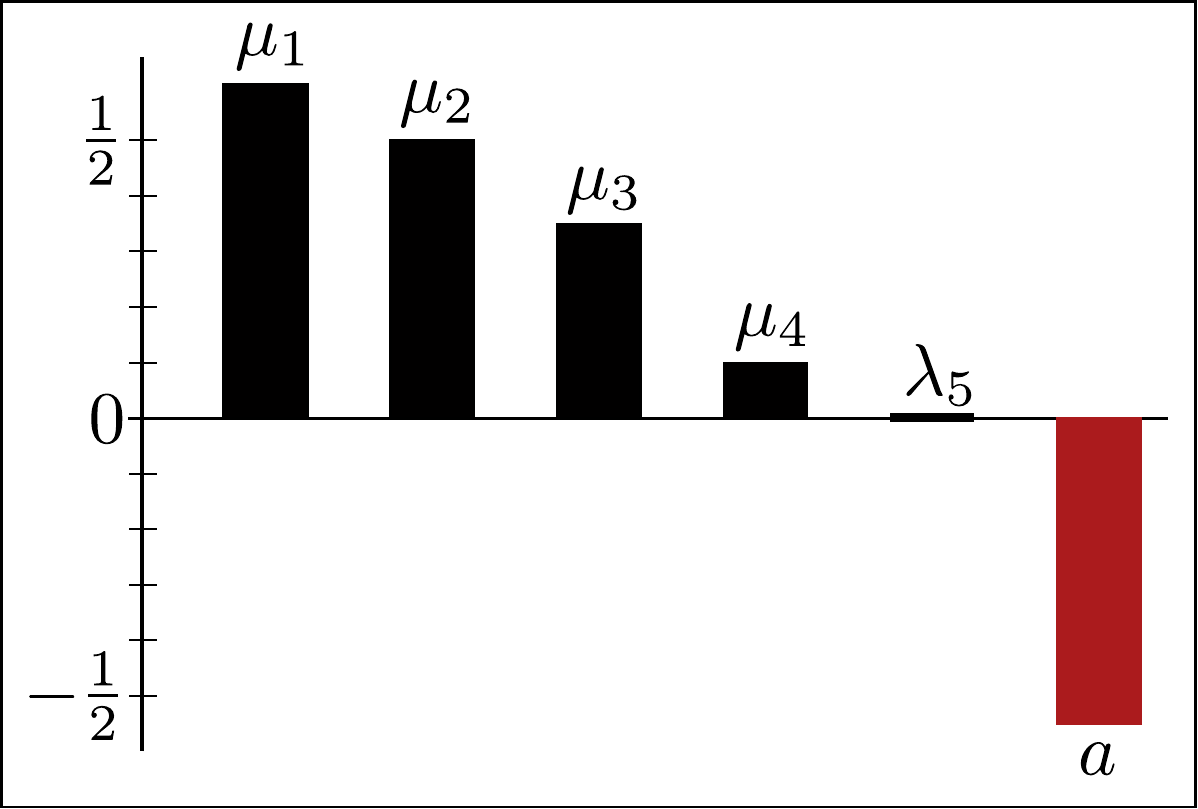}}\\
\subfloat[]{\includegraphics[width=4.1cm]{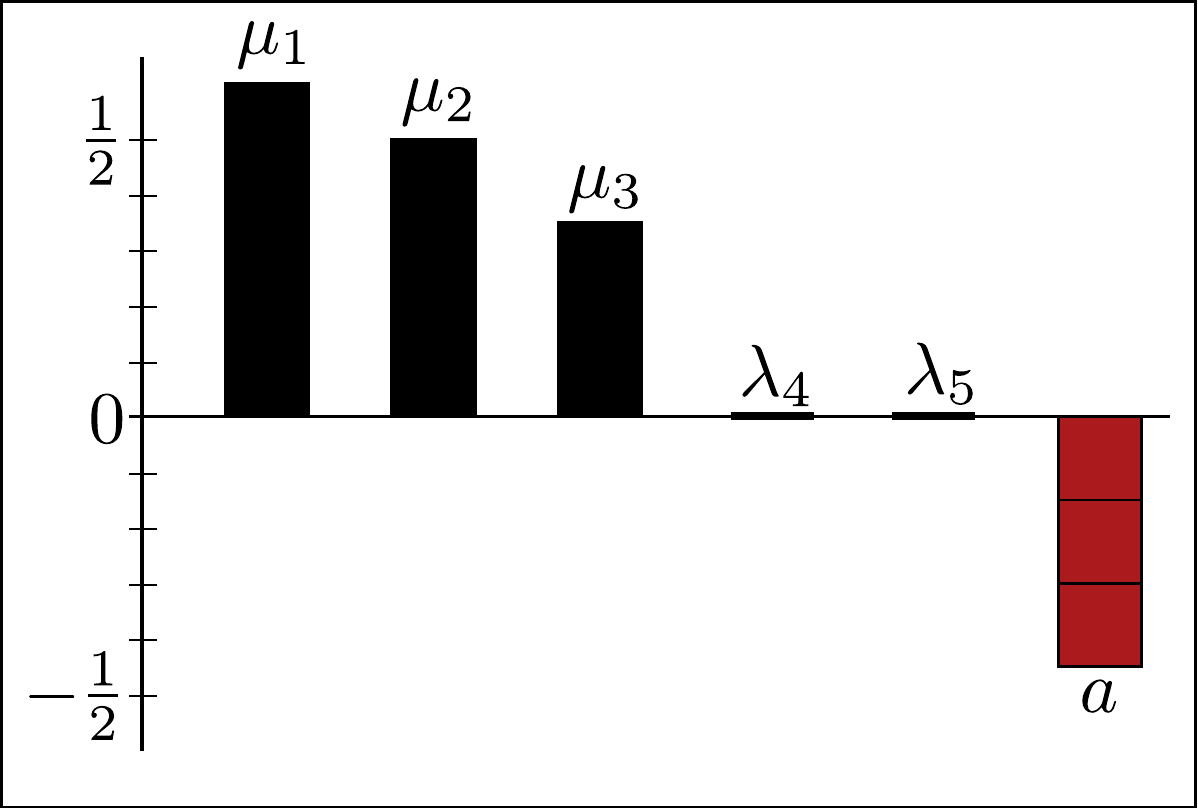}}
\ \ \ \subfloat[]{\includegraphics[width=4.1cm]{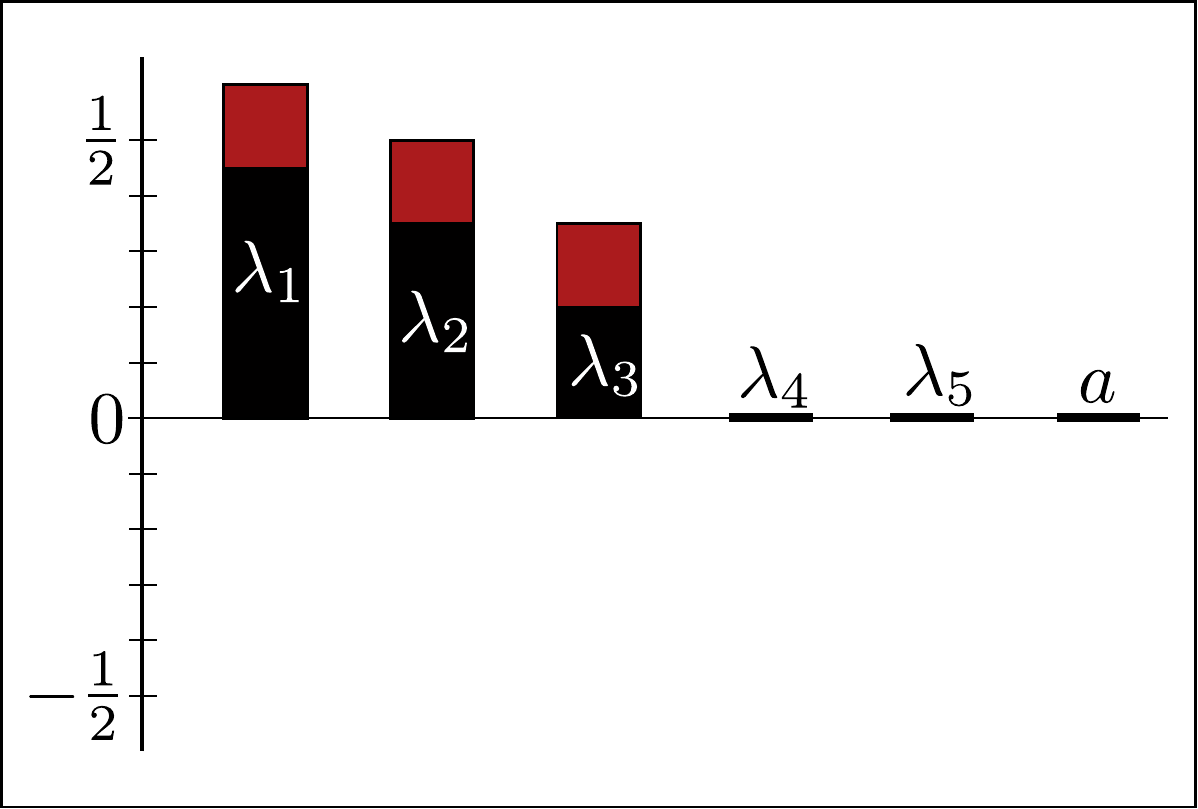}}
\caption{Example of our algorithm for subproblem 1:  (a) We start
with $\mu_1=3/5$, $\mu_2=1/2$, $\mu_3=7/20$, $\mu_4=1/10$, $\mu_5=-11/20$, and
accumulator $a=0$.  (b) Since $\mu_5+a/5$ is negative, $\lambda_5$ is set to
0 and $a$ to $-11/20$.  (c) Since $\mu_4+a/4$ is negative, $\lambda_4$ is set 
to 0 and $a=-9/20$.  (d) Finally, $\lambda_1$, $\lambda_2$, and $\lambda_3$ 
each have $a/3=-3/20$ added to the corresponding $\mu$.  The final result is
$\lambda_1=9/20$, $\lambda_2=7/20$, $\lambda_3=1/5$, and $\lambda_4=\lambda_5=0$.
}
\label{alg}
\end{figure}

Fig.~\ref{alg} works through an example of this algorithm.

\section{Efficiency of the algorithm}
\begin{figure}
\includegraphics[scale=.5]{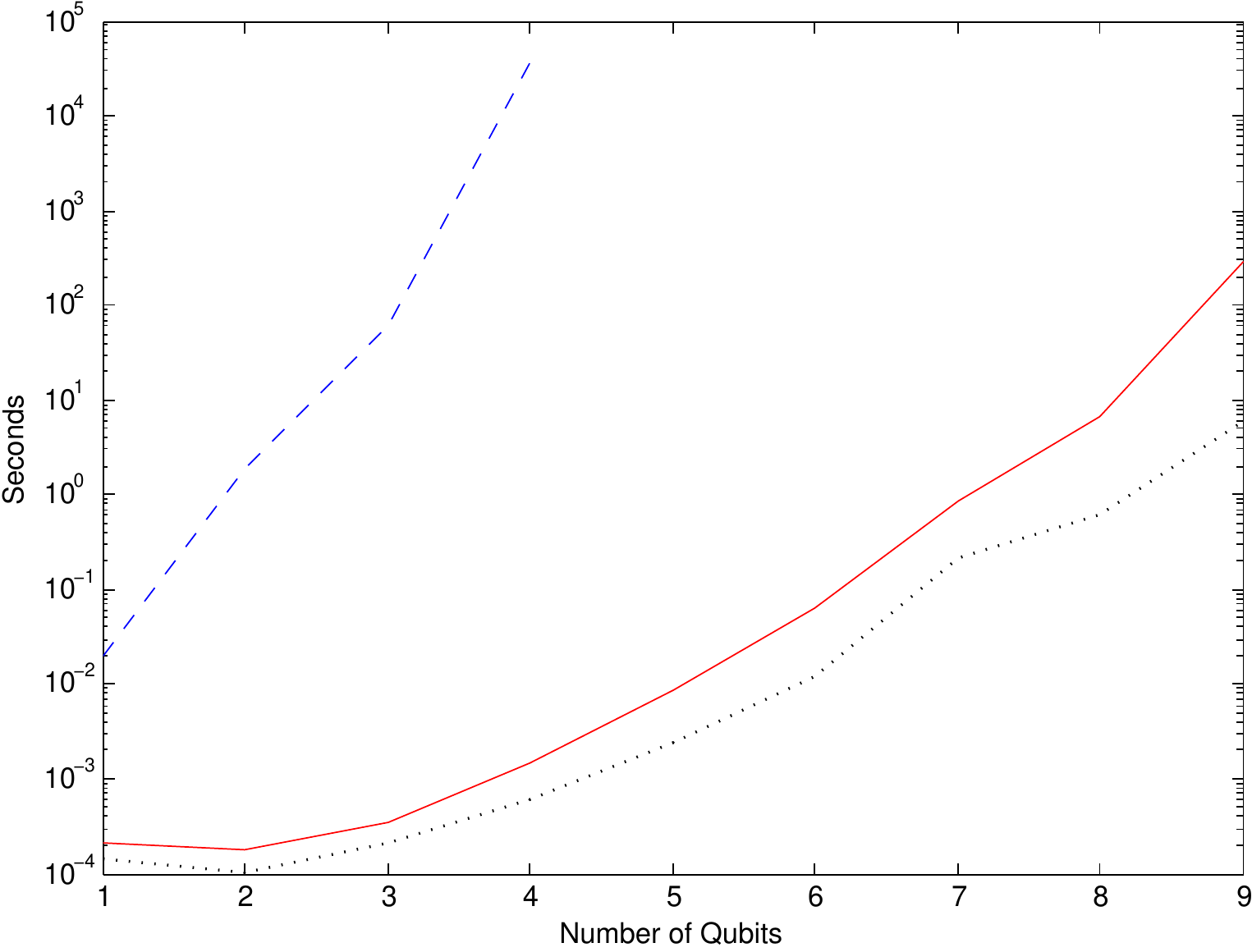}
\caption{Run time for reconstruction of random n-qubit pure states
  mixed with the identity and subjected to Gaussian noise on Pauli
  measurements.  Given a set of measurement outcomes, we generate a
  candidate matrix $\mu$ consistent with the data but with potentially
  negative eigenvalues.  We use two techniques to generate the maximum
  likelihood density matrix $\rho$.  The top, dashed line, is Matlab's
  {\bf fminsearch} used to minimize $\trace[(\mu-\rho)^2]$ directly.
  The lowest, dotted line, is our algorithm for Subproblem 1.  The
  middle, solid line, is the Subproblem 1 algorithm together with the
  basis change from the measurement outcomes to $\mu$.  All timings
  were performed in Matlab on a single core of an Intel E8400 CPU
  running at 3GHz. }
\label{figure}
\end{figure}

The slowest step is step 1, solving the eigensystem, which is $O(d^3)$
(the standard libraries such as LAPACK start and are limited by
reducing a Hermitian matrix to tridiagonal form using the Householder
method which is $O(d^3)$ \cite{householder}).  Step 2 is obviously
constant time, and step 3 and 4 together are easily seen to be $O(d)$.
Step 5 involves a choice of whether one wants the answer in the
eigenbasis of $\mu$, in which case it has already been computed, or in
some other basis, in which case it is $O(d^3)$ or better
\cite{basistransformation}.  Thus, the overall complexity is $O(d^3)$.
The actual run-time of an implementation will depend primarily on the
eigensystem solver used.  Fig. \ref{figure} compares the run time of our
algorithm with that of a traditional ML optimization.

\section{Discussion} 
Our argument that the problem of ML state reconstruction reduces to
finding the closest density matrix under the $2$-norm to a nonphysical
candidate matrix $\mu$ depended on the measurements of the expectation
values of $\trace{(\sigma_i \rho)}$ in Eq. (\ref{calL}) having the
same Gaussian variance for all $i$.  In practice, this may not be the
case.  One solution to this is to equalize the variances by simply
performing more of the noisier measurements.  For example, in circuit
quantum electrodynamics (superconducting qubits) quantum
non-demolition measurement of a Paulis are performed with $v =
1/(\Gamma \tau)$ where $\Gamma$ is the measurement rate and $\tau$ is
the measurement time.

We have focused only on state reconstruction, but it is often desired
to perform {\em process tomography} \cite{ikemikenotbook}, 
that is, to determine the quantum
input/output relation (a trace-preserving completely-positive map) implemented
by an apparatus.  Due to the Choi-Jamiolkowski isomorphism \cite{choi,jamio}
such a map can be represented by a density matrix.  Thus, the generalization
of our result to process tomography should be straightforward.

Finally, we note a connection to classical probability theory.  
If one considers the eigenvalues $\mu_i$ (some of which
may be negative) as a noisy view of a {\em probability distribution}, then
the algorithm starting from step 2 is an algorithm for finding the 
nearest proper probability distribution.  Furthermore, if the noise 
is Gaussian, this finds the maximum likelihood probability distribution.

\section{Acknowledgments} We acknowledge Jerry Chow, Antonio Corcoles,
and Matthias Steffen for valuable discussions.  We especially
appreciate Scott Glancy's discovery of an error in an earlier version
of this paper.  J.A.S. and J.M.G. were supported by the IARPA MQCO program
under contract no. W911NF-10-1-0324.


\begin{thebibliography}{99}
\bibitem{basistransformation} Basis transformation on a density matrix
  is accomplished by conjugation by unitaries, which involves matrix
  multiplication.  Matrix multiplication is $O(d^3)$ by the most
  naive method, but can be done is $O(d^{2.376})$ with a method that
  is beneficial only for impractically large $d$
  \protect\cite{Coppersmith}.  A more practical method sometimes
  employed for reasonable sized matrices (though which used with
  caution since it is less numerically stable than the naive method)
  runs in $O(d^{\log_2 7}) \approx O(d^{2.807})$  \protect\cite{Strassen}.
\end{thebibliography}
\end{document}